\documentclass{emulateapj}
\usepackage{times}

\usepackage{ifthen}

\newcommand{\Fermi}{\emph{Fermi}}
\newcommand{\FGST}{\emph{Fermi Gamma-ray Space Telescope}}
\newcommand{\gray}{$\gamma$-ray}

\newcommand{\pubbook}[6]{#1, #6, #2 #3, #4, #5}
\newcommand{\pubbooka}[6]{#1, #6, #2 #3, #4}
\newcommand{\pubbookb}[6]{#1, #6, #2 #4}
\newcommand{\pubjournal}[6]{#1 #5, #2, #3, #4}

\newcommand{\pubjournalc}[6]{#1 #5,     #4}

\newcommand{\geant}{GEANT4\ }

\newcommand{\app}{APh}

\def\IB{IGRB}

\shorttitle{Cosmic-ray induced albedo from the Oort Cloud}
\shortauthors{Moskalenko \& Porter}

\begin{document}

\title{
Isotropic \gray{} background: cosmic-ray induced albedo from debris in the solar system? 
}

\author{Igor V. Moskalenko\altaffilmark{1}}
\affil{
Hansen Experimental Physics Laboratory, 
Stanford University, Stanford, CA 94305
\email{imos@stanford.edu}}
\altaffiltext{1}{Also 
Kavli Institute for Particle Astrophysics and Cosmology,
Stanford University, Stanford, CA 94309
}

\and

\author{Troy A. Porter}
\affil{
  Santa Cruz Institute for Particle Physics,
  University of California, Santa Cruz, CA 95064
\email{tporter@scipp.ucsc.edu}}

\begin{abstract}
We calculate the \gray{} albedo due to cosmic-ray interactions with debris 
(small rocks, dust, and grains)
in the Oort Cloud.
We show that under reasonable assumptions
a significant proportion of what is called 
the ``extragalactic \gray{} background'' 
could be produced at the outer frontier of the solar system
and may be detectable by the Large Area Telescope, 
the primary instrument on the \FGST{}. 
If detected it could provide
unique direct information about the total column density of material 
in the Oort Cloud that is difficult to access by any other method. 
The same \gray{} production process 
takes place in other populations of small solar system bodies 
such
as Main Belt asteroids, Jovian and Neptunian Trojans, and 
Kuiper Belt objects.
Their detection can be used to constrain the total mass of debris in 
these systems. 
\end{abstract}


\section{Introduction}\label{intro}

Studies of the outermost region of the solar system 
started in antiquity as documented by the historical record of comet
observations \citep{Kronk1999}.
Among the oldest are Babylonian inscriptions referring to a comet of 674 B.C.
The first proven observation of the 1P/Halley
comet is recorded in the Chinese text \emph{Shih chi} 
dated 239 B.C.,
and its revisits of the inner solar system have been recorded by
subsequent generations of 
astronomers. 
However, ancient astronomers had very little knowledge about the origin
of these comets.
Only in 1950, based on observations of 
a handful of so-called long-period (LP) comets 
was the conclusion made 
that there is a huge reservoir of $\sim$$10^{11}$ comets that 
surrounds the solar system at distances larger than $\sim$$10^4$ AU 
\citep{Oort1950}.
Since that time the total number of identified LP comets has 
increased to $\sim$400 \citep{Marsden2003}.
This reservoir is now known as the Oort cloud (OC) and its outer edge 
is placed between $5\times10^4$ and $2\times10^5$ AU 
$\sim1$ pc \citep{Weissman2000,Dones2004}.
The OC population is the least explored in the solar system because
its extremely distant location from the Sun renders the detection
methods used for objects such as Kuiper Belt Objects (KBOs) 
ineffective\footnote{Here we mean
the usual optical- and infrared-based methods relying on the detection of
scattered solar radiation from the individual objects.}.

Perturbations due to the Galactic tide, passing stars, and 
giant molecular clouds
alter the orbits of objects in the OC. 
The result is that a small
fraction of these objects are injected into the inner solar system,
appearing as LP comets. 
Such comets are the only members of the OC 
population that are available for study by astrophysical methods so far.
Therefore, conclusions on the structure of the OC and
composition of its bodies are mainly based on the studies of LP 
comets. 
As the comets are largely composed of ices,
other OC bodies are assumed to have a similar composition. 
The sizes of the comet nuclei range between a few km to $\sim$50
km. 
A population of bodies below a sub-kilometer size is \emph{terra incognita}.

The OC is believed to be a remnant of the
proto-planetary disk that formed around the Sun approximately 4.6
billion years ago. 
The total mass and geometry of the OC
are determined by very indirect methods and rely 
mostly on computer-intensive simulations \citep[e.g.,][]{Dones2004}. 
The planetesimals comprising the OC initially coalesced much 
closer to the Sun, but gravitational interactions with giant planets
ejected the proto-planetary material into the outer edge of the solar
system \citep[e.g.,][]{Morbidelli2005}. 
As the accretion and collision rates in the OC are 
extremely slow, it is believed that it contains the most pristine
material left over from the epoch of planet formation.

A new method to study the populations of small solar system bodies (SSSBs -- 
such
as 
Main Belt asteroids (MBAs),
Jovian and Neptunian Trojans, and 
KBOs) 
has been recently proposed by \citet{Moskalenko2008}. 
It was shown that the 
cosmic-ray (CR) induced \gray{} flux from 
SSSBs which are large enough for the CR cascade
to fully develop in the rock (typically 
$\gtrsim 1$ meter in diameter 
-- the ``thick target'' case) 
strongly depends on the size distribution of the SSSB 
populations and may be detectable with the Large Area Telescope (LAT), the 
primary instrument on the \FGST{} (\Fermi)
(formerly the \emph{Gamma-ray Large Area Space Telescope (GLAST)}).
The \gray{} spectrum from these processes is very steep with an effective
cutoff around 1 GeV.
If detected by the LAT, it can provide unique information about the 
number of objects with sizes down 
to a few meters in each system.

For objects with diameters below $\sim 1$ meter, the 
CR cascade does not fully develop 
since the total column density of material is 
$\lesssim 1$ interaction length.
In this ``thin target'' case, the albedo \gray{s}
depend only on the total column density of material in a particular direction.
The albedo spectrum 
has a shape characteristic
of $pp$-interactions where the high energy spectral slope is similar to
that of the incident CR spectrum.
Therefore, it is possible also using \gray{} observations 
to put constraints on the total amount of debris in each system.

The outermost frontier of the solar system, the OC, has a wide distribution
on the sky.
This is different to other populations of small bodies such as the MBAs,
Jovian and Neptunian Trojans, and KBOs which are distributed near the ecliptic. 
The albedo 
\gray{s} produced by CR interactions with debris in the OC
apparently contribute to the isotropic 
\gray{} background (IGRB), which is usually assumed to be extragalactic. 
A correct estimate of the \gray{} albedo
of the OC is important for disentangling the truly extragalactic 
component. 
On the other hand, an estimate of the total column density of the 
dispersed material in the OC may shed light on the early history of the solar
system and constrain planetary evolution models. 

\section{Isotropic \gray{} background}\label{egrb}
The 
\IB\ was first discovered by the SAS-2 satellite \citep{Thompson1982} and
confirmed by EGRET \citep{Sreekumar1998}. 
The \IB\ is thought to be a superposition of all
unresolved sources of high-energy \gray{s} in the universe plus any
truly diffuse component.  
A list of the contributors to the \IB\
includes ``guaranteed'' sources such as blazars and normal galaxies
\citep{Bignami1979,Pavlidou2002}, and potential sources
such as galaxy clusters \citep{Ensslin1997}, shock waves associated
with large scale cosmological structure formation 
\citep{Loeb2000,Miniati2002}, distant \gray{} burst events 
\citep{Casanova2007}, 
pair cascades from TeV \gray{} sources and ultra-high energy CRs at
high redshifts (so-called Greisen-Zatsepin-Kuzmin cut-off).  
A consensus exists that a population of unresolved AGN certainly
contribute to the \IB\ inferred from EGRET observations; however
predictions range from 25\% up to 100\% of the \IB\
\citep{Stecker1996,Mukherjee1999,Chiang1998,Mucke2000}. 
A summary of the conventional contributors to the \IB\ can be found
in \citet{Dermer2007}.
A number of exotic sources that may contribute to the \IB\
have also been proposed: baryon-antibaryon annihilation phase after
the Big Bang, 
evaporation of primordial black holes, 
annihilation of so-called weakly interacting massive particles (WIMPs),
and strings.

The \IB\ is a
weak component which is difficult to disentangle from the intense Galactic
foreground. 
Extensive work has been done \citep{Sreekumar1998} to derive the
spectrum of the \IB\ based on EGRET data.  
A new detailed model of the Galactic diffuse 
emission \citep{Strong2004a,Moskalenko2000}
lead to a new estimate of 
the \IB\ \citep{Strong2004b} which is lower and steeper
than found by \citet{Sreekumar1998}; it is not consistent with a
power-law and shows some positive curvature, as expected for an origin
in blazars.
But, recent work has shown deficiencies in the understanding of the
\IB. 
Two more diffuse emission components originating nearby in the solar system
have been identified:
\gray{} emission due to inverse Compton scattering of solar photons by CR
electrons \citep{Moskalenko2006,Orlando2007} and a \gray{} glow
around the ecliptic due to the CR-induced \gray{} albedo of SSSBs 
\citep{Moskalenko2008}.
The former has been detected in the EGRET data at a level consistent with 
predictions \citep{Orlando2008}.
Significantly, the level of emission from this process is comparable to 
the current estimate of the \IB\ within $10^\circ$ of the Sun path on the 
sky.

\section{The Oort Cloud}\label{oc}

Although an extensive literature on the origin, population, and dynamics of 
the OC exists, 
properties of the OC population 
are derived mostly from computer simulations and 
studies of the LP and Halley-type comets 
which are thought to originate in the OC. 
In this section, we give a brief overview of the structure of the OC relevant
to the current investigation. 
For more detailed information
we refer the reader to the exhaustive books and reviews
\citep[e.g.,][and references therein]{Brandt2004,Dones2004,Fernandez2005}.

Observation of a sharp spike in the number distribution of comets at
near-zero but bound energies, representing orbits with semimajor
axes exceeding $10^4$ AU, led \citeauthor{Oort1950} to conclude that
the spike, a huge near-spherical cloud of icy bodies at 
$>2\times10^4$ AU (outer OC), 
had to be the source of the LP comets.
\citet{Hills1981} has shown that 
the apparent inner edge of the OC at a semimajor axis $a\sim(1-2)\times10^4$ AU
could be a selection effect due to the small probability of close stellar
passages capable of perturbing comets at smaller distances.
The number of comets that reside in the Hills cloud (inner OC) 
at semimajor axes of a few thousand AU could be significantly
larger than in the outer OC. 
The inner OC could also serve as a reservoir
that replenishes the outer OC stripped by an external perturber.
Close passages of stars perturbing the inner OC could result in 
so-called
``comet showers'', the extreme increases in cometary flux which lasts
for a few orbital periods after the passage. 
One such shower may be responsible
for a $\sim$2.5 My period of increased bombardment of the Earth which 
produced the large Popigai (100 km) and Chesapeake Bay (90 km) 
craters and several 
smaller craters $\sim$36 million years ago \citep{Farley1998}.

A possible clue to
the existence and the structure of the inner OC could be the 
Halley-type comets.
Dynamical simulations show that the source of these 
comets \citep{Levison2001}
is required to be a massive doughnut-shaped inner OC with median 
inclination between
10$^\circ$ and 50$^\circ$.  
Yet, there could exist a third innermost region, the 
so-called \emph{inner core} \citep{Fernandez2005},
which may span between the Scattered Disk component of the Kuiper Belt 
and the inner OC. 
The inner
core may harbor a large number of comets and mass, but does not exhibit itself 
due to the lack of very close stellar passages.
Dwarf planets 90377 Sedna and 148209 2000 CR$_{105}$ could be the
first inner core candidates. 

Dynamical models put the total number of comets with $a\ga(1-2)\times10^4$ AU
at $\sim$$10^{12}$. 
Assuming an average mass for a comet $4\times10^{16}$ g,
the total mass is estimated at $7M_\oplus$.
The more massive inner OC could harbor $\sim$$(2-13)\times10^{12}$ comets,
which yields a mass of $14-90$ $M_\oplus$.
There is no information on the total mass and number of 
sub-kilometer
size bodies.

Even closer to the Sun, beyond Neptune's orbit 
there is the Kuiper Belt. 
The KBOs 
are not uniformly distributed, with at
least three dynamically distinct populations identified: the
Classical Disk, the Scattered Disk with large eccentricities and
inclinations, and ``Plutinos'' around the 3:2  mean motion resonance
with Neptune at 39.4 AU.  
KBOs are distributed between
30 -- 100 AU  \citep[][and references therein]{Backman1995}.
A majority of the KBOs have their orbits
distributed near the ecliptic with FWHM of the order of 10$^\circ$
in ecliptic latitude \citep{Brown2001}.
The total mass of KBOs is estimated to be in the 
range $\sim$0.01--0.3 $M_\oplus$,
while the most often used value is $\sim$0.1 $M_\oplus$
\citep{Luu2002}. 
A summary of the mass and size distributions 
of KBOs, MBAs, and population of small bodies in Jovian and Neptunian Trojan
families can be found in \citet[][]{Moskalenko2008}.

The mass and size distributions for populations of 
asteroid families or other SSSBs are 
thought to be 
governed by collisional evolution and accretion. 
Under the assumptions of scaling of the
collisional response parameters and an upper cutoff in mass, the
relaxed size distributions approach a power-law $dN\propto r^{-n} dr$ \citep{Dohnanyi1969},
where $r$ is the radius of the body, and $n=3.5$ for a pure \citeauthor{Dohnanyi1969} cascade. 
In reality
different populations of SSSBs show deviations from this index.

The dust particles and grains in the solar system 
are subject to many processes which
produce a negligible effect on larger bodies, therefore
simulations of the dynamics of the 
asteroid populations 
can not tell much about the total mass of debris nor the size distribution.
The main forces acting on small particles are \citep[e.g.,][]{Grun2000}: 
gravitation, radiation pressure which is directed outward from the Sun,
the Poynting-Robertson effect which is essentially a drag reducing the 
orbital eccentricity and causing the particles to slowly spiral toward
the Sun, sublimation, collisions among dust particles, plus
interactions with the solar wind and magnetic field due to a small
electric charge carried by grains which further complicates their dynamics.
The relative effect of these processes is different across the range of
particle sizes and number densities. 
Thus, it is very difficult to simulate the distribution
of debris in the solar system.
However, 
if the column density of the debris is
large enough its distribution can be directly probed with the 
\Fermi/LAT \gray{} 
telescope.

\section{Calculations}\label{calculations}
Our previous calculations \citep{Moskalenko2008} considered bodies large 
enough to be in the thick-target case.
To calculate the CR-induced \gray{} albedo of SSSBs,
the Moon spectrum has been used as a template and scaled according to the
size distributions and densities of bodies in different populations, such as
the MBAs, Jovian and Neptunian Trojans, and KBOs.
The CR-induced \gray{} albedo spectrum of the Moon itself has been calculated
by \cite{MP2007} using a Monte Carlo code based on the \geant{} framework.
Most of the Lunar \gray{} emission comes from the thin rim with 
a steep fall off with energy, essentially cutting off above 3--4 GeV.
The central portion of the Moon disk has an even steeper 
spectrum with a cutoff above  $\sim$600 MeV.
The model calculations are in excellent agreement with the EGRET 
observations of the Moon \citep{Thompson1997,Orlando2008}.

For objects with sizes smaller than $\sim 1$ meter the CR 
cascade does
not fully develop; 
this is 
similar to CR interactions with a single nucleus
with the produced \gray{} flux scaling linearly with the total amount
of material in the interaction column.
The \gray{} spectrum in this case is harder than the thick-target 
case 
with a spectral slope close to that of the parent CR nuclei.
The \gray{} albedo of larger bodies has a soft 
spectrum which will also contribute
to the total \gray{} flux $\la$1 GeV. 
Therefore, the total albedo spectral shape below/above 1 GeV will depend on the 
relative abundances of the debris material and the larger rocks.

Production of \gray{s} in $pp$-interactions from the 
decay of neutral pions and kaons has been discussed
in many papers 
(e.g., \citealp{Stecker70,Badhwar77,StephensBadhwar81,Dermer86a,Dermer86b},
and more recently \citealp{Kamae2006} and \citealp{Kelner2006}). 
We calculate the \gray{} flux using the method described in \citet{MS1998} 
which is based on the work of 
\citet{Dermer86a,Dermer86b}.
For collisions involving
nuclei, 
the corresponding cross section is
multiplied by a factor $(A_1^{3/8}+A_2^{3/8}-1)^2$
where $A_1$ and $A_2$ are the beam and target nucleus atomic numbers
\citep{OrthBuffington76,Dermer86a}, while the energy per nucleon is the
kinematic variable. 
The accuracy of this method is enough for the current calculations.

Since oxygen is the most abundant element in the interstellar
rock and ice we calculate the production cross section 
of \gray{s} for CR protons 
and alphas interacting with an oxygen target nucleus.
The cross section is converted from nucleus$^{-1}$ to g$^{-1}$ units 
by multiplying by $N_A/15.9994$ where $N_A$ is Avogadro's number.
The total \gray{} emissivity is obtained by integrating the respective 
production
cross sections with the interstellar (unmodulated) 
incident spectrum of CR protons and alphas taken 
from \citet{MP2007}.

\begin{figure}[t]
\centerline{
\includegraphics[width=3.5in]{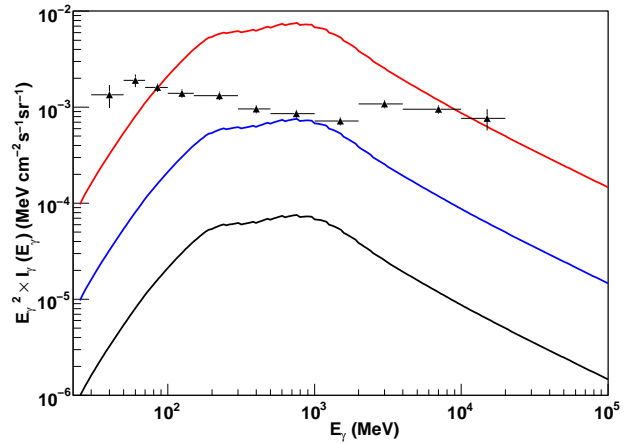}}
\caption{
Intensity of the \IB\ as derived from the EGRET data \citep{Strong2004b}. 
Curves are shown for the thin target case for different 
column densities (top to bottom): 
0.01, 0.001, 0.0001 g cm$^{-2}$.
}
\label{fig1}
\end{figure}

Figure~\ref{fig1} shows the spectrum of the \IB\ derived from 
the EGRET data \citep{Strong2004b}
together with 
the \gray{} albedo of the OC debris (thin-target case) shown for
different total column densities (top to bottom): 
0.01, 0.001, 0.0001 g cm$^{-2}$. 
The latter can be compared
with the total gas column density along the line of sight to the edge 
of the Galactic disk through the Galactic center:
$\sim$0.15 g cm$^{-2}$ assuming an average 1 H-atom cm$^{-3}$ and 
a maximum Galactocentric radius $\sim$20 kpc.


The integral \gray{} emissivity in the thin-target case can be easily estimated:

\begin{eqnarray}
S(E_\gamma > 100\ {\rm MeV}) & \approx & 6 \times10^{-2} \ {\rm g}^{-1} \ {\rm s}^{-1},\\
S(E_\gamma > 1\ {\rm GeV}) & \approx & 5 \times10^{-3} \ {\rm g}^{-1} \ {\rm s}^{-1}
\end{eqnarray}

\noindent
and the integral intensity simply following by multiplication with the 
column density, $x$, e.g., 
$I(E_\gamma > 1\ {\rm GeV}) = x S(E_\gamma > 1\ {\rm GeV})$ cm$^{-2}$ s$^{-1}$, where $x$ is in units g cm$^{-2}$.

Any assumption of the spatial distribution of debris would be highly 
speculative. 
For an order of magnitude estimate we take the simplest case 
of debris concentrated in a spherical shell with radius $d$. 
In this case
the total mass of debris is:
\begin{equation}
M_d \sim 0.47 d^2 x M_\oplus,
\label{eq1}
\end{equation}
where $d$ is the distance in AU.
It is enough to have $\sim$$50M_\oplus$ of material spread out over $4\pi$ 
at $10^3$ AU, which corresponds to $10^{-4}$ g cm$^{-2}$, to obtain a \gray{}
albedo flux within an order of magnitude of the IGRB. 
This column density corresponds 
to $\sim$$10^{-3}$ of the total column density 
along the line of sight through the 
Galactic disk. 
Since the $d^2$ dependence is very strong, much less 
material is required to produce the same albedo \gray{} flux
at a smaller effective distance and vice versa.

\section{Discussion}\label{discussion}

The total mass and distribution of SSSBs smaller than cometary 
nuclei are highly uncertain, as is the spatial distribution of debris.
Although, simulations indicate that the number of comets
and the total mass considerably increase toward the inner edge of the OC.
It is entirely possible that
$\sim$$100$ $M_\oplus$ of debris is distributed in the vast 
space of the OC. 
Since the OC has spherical and disk components, it is possible that
the OC \gray{} albedo is brighter around the ecliptic.

The minimal detectable mass in debris for other systems can also be 
estimated (Table~\ref{t1}).
For example, 
assuming that KBOs are distributed uniformly within $\pm10^\circ$ around the 
ecliptic, the fraction of the total solid angle subtended by the Kuiper Belt is 
$\Omega/4\pi=0.173$. 
The detectable mass of KBO debris can then be calculated 
using eq.~(\ref{eq1}) 
$M_d=(\Omega/4\pi) 0.47 d^2 x M_\oplus\sim 1.3\times10^{-2}M_\oplus$,
with $x=10^{-4}$ g cm$^{-2}$, and $d=40$ AU.

Jovian and Neptunian Trojans would appear as point-like sources.
The integral \gray{} flux from a point-like mass $M_d$ of debris at a 
distance $d$ AU is:
\begin{equation}
F(E_\gamma > 1\ {\rm GeV})=\frac{M_d S(E_\gamma > 1\ {\rm GeV})}{4\pi d^2} 
\sim 0.01 \frac{M_\oplus}{d^2}\ {\rm cm}^{-2} \ {\rm s}^{-1},
\end{equation}
where we used the emissivity $S(E_\gamma > 1\ {\rm GeV})$ corresponding to the 
interstellar
CR spectrum calculated earlier.
Since 1 GeV photons are produced by CR protons of $\ga$10 GeV, 
which are only slightly modulated in the heliosphere, this is a 
good approximation.

The one-year LAT integral point source sensitivity\footnote{http://www-glast.slac.stanford.edu/software/IS/glast\_lat\_performance.htm} 
is $F_{\rm 1yr}(E_\gamma > 1\ {\rm GeV})\sim 1\times10^{-9}$ cm$^{-2}$ s$^{-1}$.
This gives a detectability condition for one year of exposure:
\begin{equation}
M_d\ga 1\times10^{-7} M_\oplus d^2.
\label{eq5}
\end{equation}

The proposed method is sensitive enough to detect 
a debris mass as small as $\sim$0.1\% -- 50\% 
of the total mass of the system.
It is most sensitive for the MBA population due to its proximity,
but can also provide meaningful 
limits on more distant 
populations, such as Jovian and Neptunian Trojans, KBOs, and OC. 
A clear signature of the described process is the power-law spectrum 
of the \gray{} 
albedo ($\ga$1 GeV) which has an index similar to that of the incident
CR spectrum.
The larger photon statistics of the 
projected 5-year LAT mission will probe even 
smaller debris masses for all these systems.

\begin{deluxetable}{llll}
\tabletypesize{\footnotesize}
\tablecaption{\label{t1} Debris detection limits in different asteroid populations}
\tablecolumns{4}
\tablewidth{0pt}
\tablehead{
           & Semimajor  & Total mass, & Debris detection\\
Population & axis, AU   & $M_\oplus$ &  limit, $M_\oplus$ }
\startdata
MBAs\tablenotemark{a}
                  & 2.1--3.3 &$\sim$$6\times10^{-4}$&$\sim$$4\times10^{-7}$\\
Jovian Trojans    & 5.2      &$\sim$$1\times10^{-4}$&$\sim$$(1.8-3.8)\times10^{-6}$\\
Neptunian Trojans & 30       &$\sim$$1\times10^{-3}$&$\sim$$9\times10^{-5}$\\
KBOs\tablenotemark{b}
                  & 30--50   &$\sim$0.1             &$\sim$$1.3\times10^{-2}$\\
OC\tablenotemark{c}
                  & $>$100   &$\sim$100             &$\sim$50
\enddata
\tablenotetext{a}{Integral over the ecliptic longitude assuming average $d=2$ AU.}
\tablenotetext{b}{Using $d=40$ AU.}
\tablenotetext{c}{Using $x=10^{-4}$ g cm$^{-2}$, $d=10^3$ AU.}
\end{deluxetable}

The sensitivity and resolution of the LAT will allow it to resolve
many more individual sources, such as AGN, not resolved by EGRET 
and that contribute to current estimates of the IGRB.  
Other components of
the remaining IGRB will therefore become more important.
Understanding of the instrumental backgrounds 
within the LAT will allow discrimination of the IGRB at 10\% of the current 
level \citep{Atwood2009}.
This will allow a meaningful estimate of the foreground from
the OC and other asteroid populations to be obtained.
On the other hand, correct determination
of such a foreground will allow for more accurate determination of the truly 
extragalactic component of the diffuse emission.

\acknowledgments
I.\ V.\ M.\  and T.\ A.\ P.\ acknowledge support from a NASA
Astronomy and Physics Research and Analysis Program (APRA) grant.
T.\ A.\ P.\ acknowledges 
support from the US DOE.




\end{document}